\documentclass[11pt,twocolumn]{article}
\usepackage[utf8]{inputenc}
\usepackage[english]{babel}
\usepackage{amsmath}
\usepackage{amsfonts}
\usepackage{amssymb}
\usepackage{listings}
\usepackage{booktabs}
\usepackage{url}
\usepackage[colorlinks=true, urlcolor=blue, breaklinks=true]{hyperref}
\usepackage{geometry}
\title
{
  \textbf{Parallelizing the Factorial Space: Multi-Core OpenMP Scaling and Scalable SIMD Acceleration of the Steinhaus-Johnson-Trotter Algorithm via Dual-Lane AVX2 Execution}
}

\author{
    \textbf{Serge Melnikov} \\ 
    \small\texttt{iqfunru@gmail.com} \\
    \small\textit{Independent Researcher}
}
\date{\today}

\begin{document}

\maketitle

\begin{abstract}
This paper presents a high-performance SIMD acceleration framework for the Steinhaus-Johnson-Trotter algorithm, targeted at modern x86-64 architectures using the AVX2 instruction set. 
By exploiting a novel combinatorial space partitioning combined with single-cycle vector byte shuffling \((\texttt{\_mm256\_shuffle\_epi8}),\) our dual-lane vectorized implementation processes two independent, concurrent permutation streams within a single 256-bit YMM register under a uniform execution mask. 
Empirical evaluations demonstrate a 3$\times$ throughput increase over an optimized scalar baseline of Knuth's Algorithm P (accelerated by 3$\times$ via isolated sweeping branches) and outperform the recent Ring-Cascade algorithm by Yusheng Hu, completely avoiding store-forwarding stalls during hot loops.

To scale this engine across multi-core processors, we extend the framework into a highly concurrent environment via OpenMP using a localized mathematical state decoder and macro-period loop scheduling. 
The parallel performance is shown to scale strictly and linearly with the number of active physical processor cores (\(\text{Speedup}(M) \approx M\)) due to a lock-free thread-local accumulation pipeline that eliminates false sharing. 
On a 6-core processor, the multi-threaded engine delivers a \(\mathbf{5.3\times}\) throughput gain for $n=14$ (\(\mathbf{3.25}\) billion CPU cycles) and processes the massive $n=15$ space in just \(16.0\) seconds, yielding a \(\mathbf{6.1\times}\) speedup over the sequential vector baseline, while Hyper-Threading virtual cores yield zero additional throughput due to physical SIMD port saturation.
\end{abstract}

\section{Introduction}
We begin by defining a natural recurrence relations algorithm to generate permutations. Let the elements of a permutation be a set of integers $S = \{0, 1, 2, \dots, n-1\}$, where $n$ denotes the total number of elements. The initial permutation of the elements $n$ is defined in ascending order as:
\begin{equation}
\pi_0 = (0, 1, 2, \dots, n-1)
\end{equation}

For the base case of a single element ($n=1$), the sequence contains only one permutation: $(0)$. To generate all permutations for $n=2$, we duplicate the base sequence and systematically insert the element $1$ at all possible positions via a staircase insertion cascade:
\begin{equation*}
\begin{matrix} 0 & 1 \\ 1 & 0 \end{matrix}
\end{equation*}

To extend this to a 3-element set ($n=3$), each 2-element permutation is duplicated 3 times. The element $2$ is then systematically interleaved through all valid positions in a bidirectional staircase manner within each 2-element sub-permutation:
\begin{equation*}
\begin{matrix} 0 & 1 & 2 \\ 0 & 2 & 1 \\ 2 & 0 & 1 \\ 2 & 1 & 0 \\ 1 & 2 & 0 \\ 1 & 0 & 2 \end{matrix}
\end{equation*}

Similarly, to construct the complete set of permutations for the elements $n+1$, each permutation of order $n$ is repeated $n+1$ times, followed by the application of the staircase insertion pattern for the element $n$. We define this specific sequential output order as the \textit{canonical permutation order}. Furthermore, we index (rank) all generated permutations of order $n$ within the range $[0, n!-1]$. Below is the complete sequence of 3-element permutations listed alongside their respective canonical ranks:
\begin{equation*}
\begin{matrix} 
0: & 012 & \quad & 3: & 210 \\
1: & 021 & \quad & 4: & 120 \\
2: & 201 & \quad & 5: & 102 
\end{matrix}
\end{equation*}

We establish a \textit{hierarchy of seniority} to rank elements within a permutation: element $n-1$ is designated as the most junior (least significant), followed by element $n-2$, up to element $0$, which is designated as the most senior (top level). Empirically, more junior elements exhibit higher mobility, continuously shifting positions by swapping with their immediate left or right neighbors. For instance, element $1$ performs only a single transposition, whereas element $0$ remains stationary throughout the entire generation process.

We distinguish between \textit{active} and \textit{passive} element displacement. In the permutations ranked 1 and 2 above, element $2$ actively shifts to the left, while the elements it bypasses undergo a passive rightward displacement. During an active transposition, an element always swaps positions with a more senior element. Each element (except 0) executes directed transpositions within its respective sub-permutation (where it acts as the local maximum) until it encounters either a more junior element or the structural boundary of the global permutation. Upon hitting this boundary, the element stalls its displacement for the current step, inverts its direction of movement, and yields the execution token to a more senior element $m$. During any transposition step of a more senior element $m$, all elements more junior than $m$ are positioned strictly at the boundaries of their sub-permutations.

High-performance permutation generation plays a critical role in cryptography and graph theory algorithms. This raises a fundamental question: how can we generate all permutations in canonical order iteratively, eliminating the overhead of recursive calls? This problem is efficiently addressed by the non-recursive Steinhaus-Johnson-Trotter (SJT) algorithm \cite{johnson1963, trotter1962}.

\section{Optimized Algorithm P}
Algorithm P, as detailed by Knuth \cite{knuth2011}, represents a highly optimized iterative variant of the SJT algorithm. As explicitly noted in \cite{knuth2011}, its performance can be further accelerated by isolating the bidirectional staircase insertion path of the most junior element into a dedicated, branch-optimized execution path. Empirically, this microarchitectural separation yields an approximate threefold ($\approx 3\times$) speedup. Additional theoretical foundations of Algorithm P are documented in \cite{arcus2014}. We present an optimized scalar implementation derived from the description in \cite{knuth2011}, modified to enforce the canonical sequence and utilize zero-based array indexing. The complete source code for this algorithm is provided in the external file \texttt{p\_opt.c}.

It is critical to note that the conventional consensus established by Sedgewick and Knuth—claiming that Heap's algorithm inherently outperforms Algorithm P—is obsolete under modern architectural paradigms. On contemporary deeply pipelined superscalar processors, even a non-recursive, iterative implementation of Heap's algorithm executes approximately three times as slowly as the optimized Algorithm P. This performance degradation occurs because the permutation sequence generated by Heap's algorithm induces severe instruction pipeline stalls due to frequent branch mispredictions and irregular data access patterns.

In addition to traditional generation methods like the Johnson-Trotter-Steinhaus algorithm or Heap's method, alternative high-speed approaches have recently emerged. Specifically, the scalar Ring-Cascade Permutation Algorithm (Position-Pro variant) by Hu \cite{hu2020} achieves exceptional throughput without the structural constraints of hardware vectorization, utilizing a completely different combinatorial logic. While Hu's algorithm provides an unconstrained scalar alternative, our optimization efforts focus on enhancing...

\section{Microarchitectural Optimization Techniques}
\label{sec:optimization_techniques}
We propose two architectural and combinatorial optimization mechanisms within the SJT framework.

\subsection{Vectorized Staircase Interleaving}
The core operational principle relies on the fact that once a valid permutation of elements $n$ is constructed (e.g. $(0, 1, 2, 3)$), it can serve as a base tensor to rapidly derive subsequent permutations $n+1$ by sweeping the element $n$ across all available slots through a staircase insertion cascade. In a standard scalar execution environment, this technique counterintuitively degrades performance due to increased L1 data cache write-traffic. However, when deployed within a SIMD environment utilizing 256-bit YMM registers, this mechanism yields a substantial performance dividend. Due to register width constraints, this specific optimization bounds the permutation length to a maximum of $n \le 16$ elements.

\subsection{Exploitation of the Half-Space Reflection Invariant}
This approach leverages a fundamental combinatorial invariant of the canonical permutation output order. In this sequence, each element $i$ executes exactly $i! \cdot i$ transpositions. Notably, the element $1$ undergoes exactly one structural shift during the transition from the permutation ranked $\frac{n!}{2}-1$ to the permutation ranked $\frac{n!}{2}$. Consequently, within the first half of the generated space ($[0, \frac{n!}{2}-1]$), the element $0$ strictly precedes the element $1$, whereas in the remaining half ($[\frac{n!}{2}, n!-1]$), the element $1$ strictly precedes $0$. 

This structural asymmetry guarantees that no two permutations within the same half-space are mirror images (reversals) of one another. Therefore, it is mathematically sufficient to compute only the initial half-space. The subsequent half-space can be dynamically reconstructed by the calling subroutine via a zero-overhead right-to-left array reversal. By eliminating the execution steps required to generate the redundant half-space, this mechanism cuts the algorithmic generation workload exactly in half.

\section{Sequential Vectorized Implementation via AVX2/YMM}
Within the vectorized execution paradigm, the maximum permutation length is structurally bounded to $n \le 16$. This constraint is dictated by the architectural parameters of the x86\_64 SIMD registers, where each 128-bit lane of a 256-bit YMM register comprises exactly 16 bytes. Furthermore, the vectorized routines presented below output the permutations in a modified non-canonical sequence due to the nature of parallel lane scheduling.

We propose a high-performance vector implementation that generates two concurrent permutations within a single YMM register per single-cycle core retirement. This is achieved by utilizing the byte-shuffling intrinsic \texttt{\_mm256\_shuffle\_epi8} (the \texttt{VPSHUFB} instruction). Specifically, the lower 128-bit lane of the YMM register concurrently computes permutations mapped to the ranking interval $[0, \frac{n!}{4}-1]$, while the upper 128-bit lane simultaneously evaluates the adjacent combinatorial space from $[\frac{n!}{4}, \frac{n!}{2}-1]$.

This architecture successfully integrates both optimization mechanisms described in Section \ref{sec:optimization_techniques}. For rigorous microarchitectural benchmarking, this standalone routine isolates the generation loop by omitting the overhead of passing data to an external caller, establishing a direct baseline comparison with our optimized scalar Algorithm P.

In raw clock cycles, this vectorized implementation demonstrates a 6$\times$ execution speedup over the optimized version of Algorithm P. However, because our kernel natively computes only the primary half-space of the factorial domain, the normalized application-level throughput dividend scales to an exact 3$\times$ performance increase.

To prevent the aggressive dead-code elimination (DCE) passes of the GCC compiler from optimizing away the generation loop, we implemented a lightweight checksum verification kernel executed at runtime termination. While this validation step introduces a minor instruction-count penalty, it guarantees benchmark integrity. To enable standard console output for verification, the user can uncomment the compilation directive:

\texttt{// \#define HALF\_PRINT\_ENABLED}

The complete source code for this benchmarking routine is provided in the external file \texttt{ymm\_final\_en.c}.

Additionally, we provide a production-ready variant of the algorithm suitable for practical applications. In this implementation, each vectorized permutation matrix is passed to a user-defined callback function. For demonstration purposes, this callback executes a bidirectional sweep, printing each tensor in both direct and inverted (reversed) order. Permutation output can be activated via the same directive:

\texttt{// \#define HALF\_PRINT\_ENABLED}

The source code for this callback-integrated variant is provided in the external file \texttt{ymm\_final1\_en.c}.

\section{Multi-Threaded Parallel Implementation via OpenMP}

The concurrency architecture of the parallel permutation engine mirrors the internal lane-splitting logic of the sequential baseline, extending the micro-steps into high-performance multi-core execution via the OpenMP framework. 
Instead of repeatedly evaluating high-overhead global combinatoric offsets or recalculating factorials within hot paths, the parallel engine allocates long-lived worker threads that invoke a highly optimized state decoder, \texttt{StructuralInitState}, strictly once upon thread initialization.

The total invariant combinatorial subspace $\frac{n!}{4}$ is seamlessly mapped across the execution pipeline by slicing the space into explicit macro-periods, bounded dynamically to guarantee that the fast-sweeping element $n-1$ always perfectly aligns to the rightmost boundary during initial worker sweeps. 
A lightweight, lock-free register accumulation topology is applied on the thread stack, allowing core execution units to modify independent memory lines within their respective L1/L2 caches without triggering false sharing or pipeline lock stalls. 

Empirical hardware verification indicates that the parallel processing throughput scales strictly, near-perfectly, and linearly with the exact number of active physical processor cores allocated to the application ($\text{Speedup}(M) \approx M$). 
Simultaneous Multi-Threading (SMT / Hyper-Threading) virtual contexts yield 0\% performance additions, providing an empirical validation that a single vectorized streaming context completely saturates the physical shuffle and shift ports of the underlying SIMD execution units.

To accommodate multi-threaded workloads, the operational syntax of the cross-platform command-line execution was extended to incorporate a secondary thread-allocation parameter. 
The sequential execution format:
\begin{verbatim}
  ./ymm_final_en <n>
\end{verbatim}
is thus superseded by the concurrent multi\-threaded execution token structure:
\begin{verbatim}
  ./ymm_final_en_mt <n> <threads>
\end{verbatim}
where \texttt{<n>} represents the permutation target bound ($6 \le n \le 16$) and \texttt{<threads>} designates the active thread allocation pool (e.g., \texttt{t5} or \texttt{5}). 
Standard OS terminal redirection operations remain fully compliant under concurrent execution, allowing the thread-safe output stream to be buffered cleanly into physical storage files:
\begin{verbatim}
  ./ymm_final_en_mt_cb 10 t7 > out.txt
\end{verbatim}

\section{Experimental Setup}
Empirical validation was conducted on a mini-PC node powered by an Intel Core i7-8850H CPU operating at a sustained frequency of 4.2 GHz (supporting instructions up to AVX2). The memory subsystem consisted of DDR4 RAM clocked at 2.67 GHz running under a 64-bit Windows 10 OS.

The source code was compiled using GCC version 15.2.0 with strict performance-oriented optimization flags enabled: \texttt{-O3 -march=native -s -Wall}. To prevent out-of-order execution leakage and ensure microarchitectural state isolation, we utilized \texttt{LFENCE} serializing instructions around the \texttt{RDTSC} hardware cycle counter, preceded by a strict core-warmup phase to stabilize CPU frequency scaling. All developed source code maintains cross-compiler compatibility under MSVC and Clang.

\section{Remarks on Further Optimizations}

It is anticipated that configuring the permutation order $n$ as a compile-time constant will yield an incremental performance dividend. 
Furthermore, an external calling routine can ingest the generated permutation tensors directly within the YMM registers, eliminating the store-to-load forwarding overhead. 
The proposed execution framework scales orthogonally to modern 512-bit vector extensions (AVX-512/ZMM) and data-parallel multithreaded topologies. 
With the introduction of the OpenMP framework, the algorithm is no longer bound by individual core frequency limits, enabling arbitrary performance scaling that directly matches the physical core count of any upcoming server-grade x86\_64 processor architecture.

\section{Conclusion}
In this paper, we presented a highly scalable parallel framework that accelerates the Steinhaus-Johnson-Trotter permutation generation algorithm on modern multi-core x86\_64 processing nodes. 
By orchestrating dual-lane SIMD stream parallelization inside a 256-bit AVX2 register mask and introducing a decentralized, lock-free thread-local initialization topology using OpenMP, the implementation completely eliminates runtime lock contention and memory bus saturation bottlenecks. 
Experimental hardware verification proves that the execution throughput scales strictly and linearly in direct proportion to the number of active physical processor cores ($\text{Speedup}(M) \approx M$). 
The framework delivers unprecedented combinatorial execution speeds, exhausting the massive $15!$ factorial space in a mere $16$ seconds on a 6-core Coffee Lake mobile CPU and demonstrating cross-platform deployment compliance across enterprise high-performance computing environments.

\section*{Code Availability}

The complete, cross-platform C source code implementing the baseline sequential and multi-threaded parallel algorithms is openly available on GitHub at
\url{https://github.com/Serge-Melnikov/parallel-avx2-permutation-generator}. 
The repository is structured into two main operational directories to maintain benchmarking isolation:
\begin{enumerate}
  \item \textbf{The \texttt{sequential/} Directory:} Contains the accelerated scalar implementation of Knuth's Algorithm P (\texttt{p\_opt\_en.c}), the single-threaded baseline vector engine for raw hardware benchmarking (\texttt{ymm\_final\_en.c}), and the single-threaded vector generator featuring a user-defined processing interface (\texttt{ymm\_final1\_en.c}).
  \item \textbf{The \texttt{parallel/} Directory:} Contains the high-performance multi-threaded vector engine designed for idle hardware core benchmarks (\texttt{ymm\_final\_en\_mt.c}) and the practical multi-threaded production generator featuring a thread-safe user callback execution pipeline (\texttt{ymm\_final\_en\_mt\_cb.c}).
\end{enumerate}

\end{document}